\documentclass[aps,prl,twocolumn,nopacs,superscriptaddress]{revtex4}

\usepackage{graphicx}  
\usepackage{dcolumn}   
\usepackage{bm}        
\usepackage{amssymb}   
\usepackage{amsmath}
\usepackage{units}
\usepackage[ansinew]{inputenc}
\usepackage[dvipsnames]{xcolor}

\usepackage[type1]{libertine}                                        
\usepackage{textcomp}
\usepackage[scaled=.85]{beramono}
\usepackage[libertine,cmintegrals,cmbraces,vvarbb,slantedGreek]{newtxmath}
\usepackage[scr=boondoxo]{mathalfa}
\usepackage{bm}
\usepackage[lf]{carlito}

\usepackage{braket}
\usepackage{bm}

\makeatletter
\DeclareRobustCommand{\cev}[1]{%
  \mathpalette\do@cev{#1}%
}
\newcommand{\do@cev}[2]{%
  \fix@cev{#1}{+}%
  \reflectbox{$\m@th#1\vec{\reflectbox{$\fix@cev{#1}{-}\m@th#1#2\fix@cev{#1}{+}$}}$}%
  \fix@cev{#1}{-}%
}
\newcommand{\fix@cev}[2]{%
  \ifx#1\displaystyle
    \mkern#23mu
  \else
    \ifx#1\textstyle
      \mkern#23mu
    \else
      \ifx#1\scriptstyle
        \mkern#22mu
      \else
        \mkern#22mu
      \fi
    \fi
  \fi
}
\makeatother

\renewcommand{\Re}{\text{Re}}

\begin{document}

\title{Single Channel Josephson Effect in a High Transmission Atomic Contact}

\author{Jacob Senkpiel}
\affiliation{Max-Planck-Institut f\"ur Festk\"orperforschung, Heisenbergstraße 1,
70569 Stuttgart, Germany}
\author{Simon Dambach}
\affiliation{Institut für Komplexe Quantensysteme and IQST, Universität Ulm, Albert-Einstein-Allee 11, 89069 Ulm, Germany}
\author{Markus Etzkorn}
\affiliation{Max-Planck-Institut f\"ur Festk\"orperforschung, Heisenbergstraße 1,
70569 Stuttgart, Germany}
\author{Robert Drost}
\affiliation{Max-Planck-Institut f\"ur Festk\"orperforschung, Heisenbergstraße 1,
70569 Stuttgart, Germany}
\author{Ciprian Padurariu}
\affiliation{Institut für Komplexe Quantensysteme and IQST, Universität Ulm, Albert-Einstein-Allee 11, 89069 Ulm, Germany}
\author{Björn Kubala}
\affiliation{Institut für Komplexe Quantensysteme and IQST, Universität Ulm, Albert-Einstein-Allee 11, 89069 Ulm, Germany}
\author{Wolfgang Belzig}
\affiliation{Fachbereich Physik, Universität Konstanz, 78457 Konstanz, Germany}
\author{Alfredo Levy Yeyati}
\affiliation{Departamento de F\'{\i}sica Te\'orica de la Materia Condensada, Condensed Matter Physics Center (IFIMAC), and Instituto Nicolás Cabrera, Universidad Autónoma de Madrid, 28049 Madrid, Spain}
\author{Juan Carlos Cuevas}
\affiliation{Departamento de F\'{\i}sica Te\'orica de la Materia Condensada, Condensed Matter Physics Center (IFIMAC), and Instituto Nicolás Cabrera, Universidad Autónoma de Madrid, 28049 Madrid, Spain}
\author{Joachim Ankerhold}
\affiliation{Institut für Komplexe Quantensysteme and IQST, Universität Ulm, Albert-Einstein-Allee 11, 89069 Ulm, Germany}
\author{Christian R. Ast}
\email[Corresponding author; electronic address:\ ]{c.ast@fkf.mpg.de}
\affiliation{Max-Planck-Institut f\"ur Festk\"orperforschung, Heisenbergstraße 1,
70569 Stuttgart, Germany}
\author{Klaus Kern}
\affiliation{Max-Planck-Institut f\"ur Festk\"orperforschung, Heisenbergstraße 1,
70569 Stuttgart, Germany}
\affiliation{Institut de Physique, Ecole Polytechnique Fédérale de Lausanne, 1015 Lausanne, Switzerland}

\date{\today}

\begin{abstract}
The Josephson effect in scanning tunneling microscopy (STM) is an excellent tool to probe the properties of the superconducting order parameter on a local scale through the Ambegaokar-Baratoff (AB) relation. Using single atomic contacts created by means of atom manipulation, we demonstrate that in the extreme case of a single transport channel through the atomic junction modifications of the current-phase relation lead to significant deviations from the linear AB formula relating the critical current to the involved gap parameters. Using the full current-phase relation for arbitrary channel transmission, we model the Josephson effect in the dynamical Coulomb blockade regime because the charging energy of the junction capacitance cannot be neglected. We find excellent agreement with the experimental data. Projecting the current-phase relation onto the charge transfer operator shows that at high transmission multiple Cooper pair tunneling may occur. These deviations become non-negligible in Josephson-STM, for example, when scanning across single adatoms.
\end{abstract}

\maketitle


Control of electronic properties in quantum-coherent nanostructures such as Josephson junctions is difficult to achieve as it requires deterministic structure design at the atomic scale. Without atomic scale design the conductance of identically prepared nanostructures exhibits fluctuations of the order of the conductance quantum $G_0=2e^2/h$. High level control has been achieved using atomic break junctions to realize few channels highly transparent Josephson atomic point contacts (JAPC) \cite{muller_experimental_1992,scheer_signature_1998,scheer_conduction_1997,chauvin_crossover_2007,della_rocca_measurement_2007}. The highlight of JAPCs is that they can be tuned to the regime where electronic transport is dominated by a single transport channel with large, nearly reflectionless, transmission. As a result, the current-phase relation of the junction becomes non-sinusoidal and multiple Cooper pair processes may occur. At the same time, the excitation spectrum of Andreev levels carrying the Josephson current consists of a single Andreev bound state (ABS) that is well separated from other ABS and from the continuum of states above the gap. Thus, the maximum supercurrent carried by a JAPC, i.\ e.\ the critical current $I_C$, does not only depend on the superconducting gap parameters in the two leads, but also on the details of the tunneling conductance \cite{ambegaokar_tunneling_1963,beenakker_universal_1991}, i.\ e.\ the number of transport channels and their transparency. This scenario has been used to study experimentally the transition from coherent Josephson transport to the regime of multiple Andreev reflections (MARs) and also to reveal for the first time coherent ABS dynamics \cite{chauvin_crossover_2007}. In these and previous studies, the superconducting phase difference behaved as a classical variable, its quantum fluctuations being negligible. Equivalently, charge quantization and charging effects could be neglected such that the Josephson current was fully determined by the classical dynamics of the phase.

Design at the atomic scale can be perfectioned using a scanning tunneling microscope (STM) through direct atomic manipulation with more control and reproducibility than in break junctions. However, a downside in the STM is the limited design flexibility concerning the influence of the environment. This implies non-negligible charging effects and quantum fluctuations of the phase \cite{ingold_cooper-pair_1994,jack_nanoscale_2015,ast_sensing_2016}. Still, thermal fluctuations can be reduced by operating in the low mK regime \cite{jack_quantum_2017}. In this new scenario, the effect of the electromagnetic environment seen by the junction leads to dynamical Coulomb blockade (DCB) type physics \cite{devoret_effect_1990,averin_incoherent_1990}, which has remained largely unexplored until recently as it requires both significant charging effects as well as a high transparency channel in order to be visible.

Here, we demonstrate in an STM single channel Josephson tunneling in the presence of DCB up to very high conductances $>0.9G_0$. We build a single atom contact by placing a single aluminium atom onto an Al(100) surface and approaching it with an atomically sharp tip made of polycrystalline aluminum. Operating at a base temperature of 15\,mK, we ensure that both tip and sample are superconducting ($T^\text{Al}_\text{C} = 1.2\,$K). We obtain a JAPC that features a single Josephson channel where the tip-sample positioning offers unprecedented reproducibility of the channel transmission coefficient, from below 0.1 to above 0.95, with all other channels having lower transmission by at least one order of magnitude. The set of transport channels with their transmission, the so called mesoscopic pin code, is extracted from measurements of the current-voltage curves in the MAR regime, using a well established technique \cite{cuevas_hamiltonian_1996,scheer_conduction_1997}. The interplay between quantum fluctuations of the phase and the fluctuations due to the electromagnetic environment is most prominent in the Josephson peak forming in the lower voltage portion of the current-voltage curve. At low transmission, the Josephson effect is well modeled using the Ambegaokar-Baratoff (AB) formula for the Josephson energy in the tunnel limit augmented by a description of the environmental interaction using $P(E)$-theory \cite{ingold_cooper-pair_1994,jack_critical_2016,randeria_scanning_2016}. However, at transmissions exceeding 0.1 the results deviate significantly from the AB formula and a general theoretical model in this regime is currently lacking. Here, we provide a simple theoretical picture where Cooper pair tunneling occurs by incoherent transfer of single and possibly multiple Cooper pairs with rates calculated using $P(E)$-theory. This type of coherence loss between tunneling events is reminiscent of DCB physics, but is in contrast to the DCB regime of conventional Josephson junctions where tunneling occurs by incoherent single Cooper pairs. The theoretical model fits very well with the measured data for voltages below the threshold where MAR processes become relevant. Thus, our measurements provide an important step towards understanding DCB physics in the single channel Josephson regime and may inspire future theories on the transition between Josephson and MAR regimes in the presence of significant quantum fluctuations of the phase. The results also provide a better understanding of the intricacies of the Josephson effect as a local probe of the superconducting order parameter \cite{hamidian_detection_2016,randeria_scanning_2016,jack_critical_2016}.


\begin{figure}
\centerline{\includegraphics[width = \columnwidth]{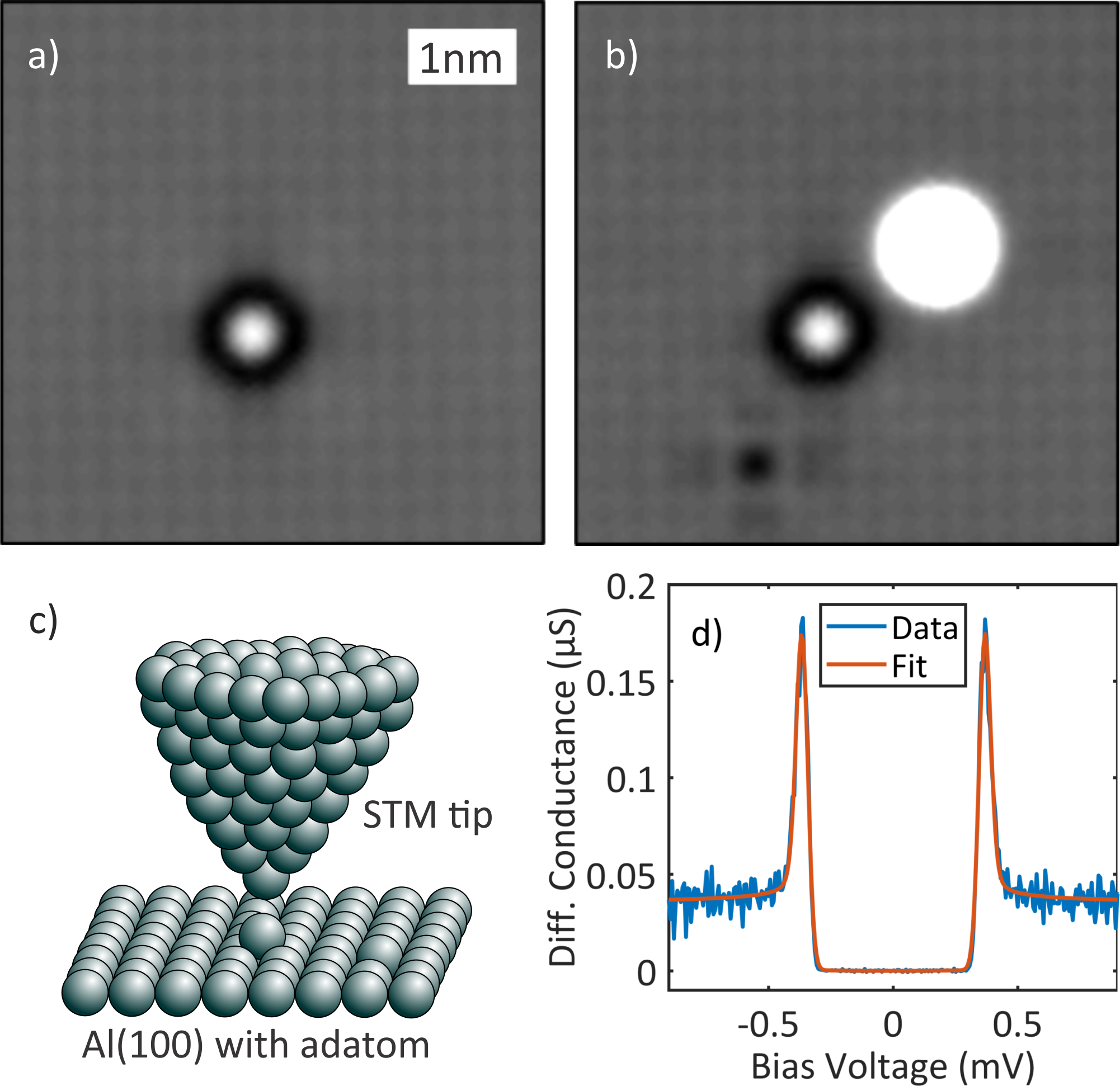}}
\caption{a) Topographic image of the Al(100) surface with an adsorbed foreign atom as a reference (white protrusion with black halo) before atomic manipulation. b) An Al atom has been pulled from the surface (black depression on the lower left) and placed on to the surface again (white protrusion on the upper right). The contrast has been adjusted to display the details of the lattice corrugation, such that the adatom appears completely white. c) Schematic of the tunnel junction. The tip of the scanning tunneling microscope is directly over the Al adatom creating a contact between two single atoms. d) Fit of a quasiparticle differential conductance spectrum at a conductance setpoint of 36\,nS, where no Josephson effect and no Andreev reflections are observed. } \label{fig:topo}
\end{figure}

We build a single-atom junction by pulling an aluminum atom with the aluminum tip from the atomically flat Al(100) surface (see Fig.\ \ref{fig:topo}(a)) and placing it on the surface again, which is shown in Fig.\ \ref{fig:topo}(b). The black depression at the lower left part of the image in (b) is the vacancy of the missing Al atom, which now appears as the rather large white protrusion (due to the image contrast) to the right of the center. This constitutes a reproducible way to create single-atom junctions with the STM as shown schematically in Fig.\ \ref{fig:topo}(c). As has been shown before, a single-atom contact does not necessarily constitute a single channel contact \cite{scheer_signature_1998,scheer_conduction_1997,cuevas_microscopic_1998}. Most atoms have more than one valence orbital that is available for electron transport.

\begin{figure}
\centerline{\includegraphics[width = \columnwidth]{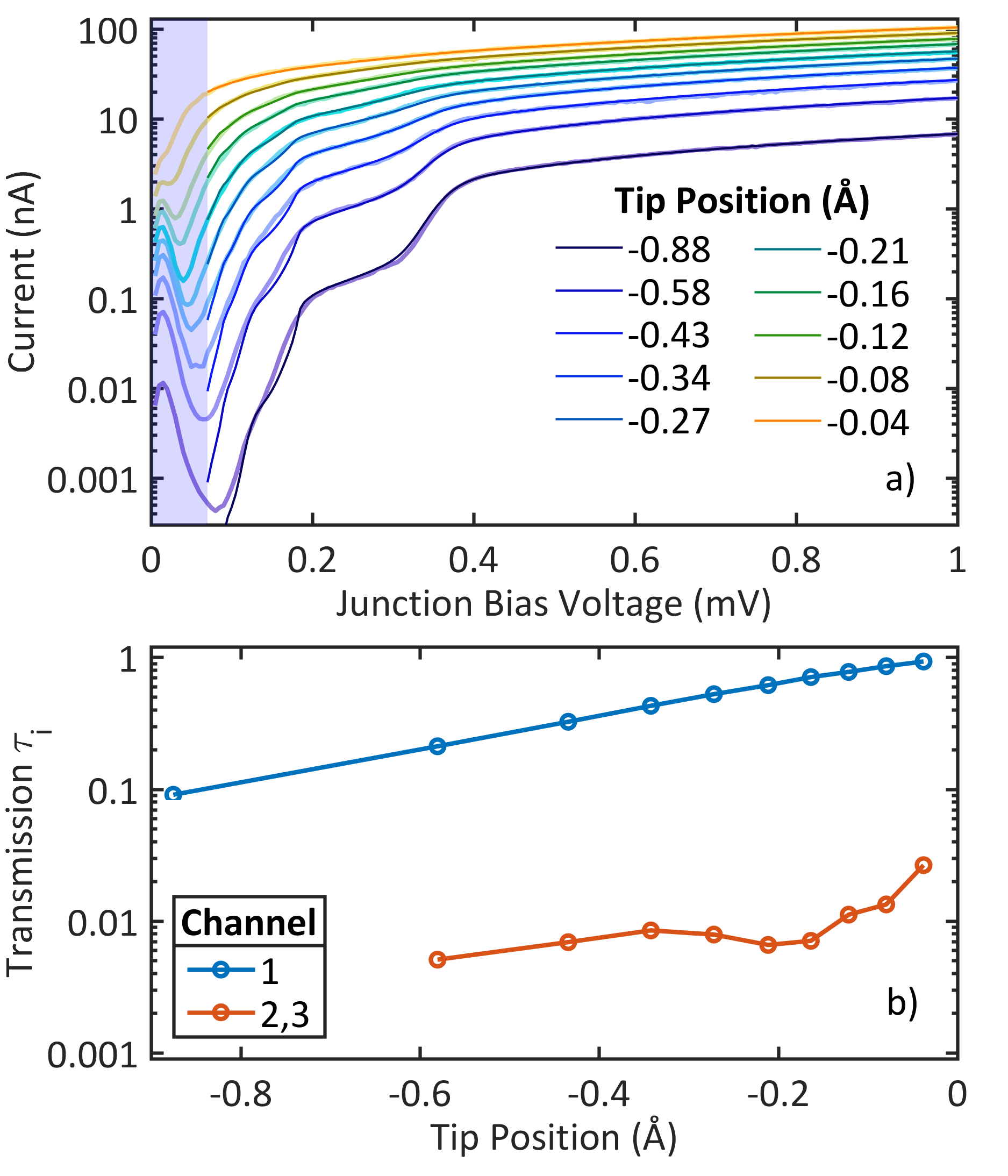}}
\caption{a) Fits (thin lines) of the Andreev reflection data (thick lines) as function of tip-sample distance, i.\ e.\ vertical tip position above the adatom on the surface (the origin of the length scale is at $G_\text{N}=G_0$). in order to extract the number of channels and their transmission. For symmetry reasons and based on the valence of Al, we have assumed three channels. The blue shaded area is the energy range for the analysis of the Josephson spectra in Fig.\ \ref{fig:jospanels}. b) The three channels and their transmission for the different spectra in a) as a function of tip-sample distance. The second and the third channel have equal transmission.} \label{fig:mars}
\end{figure}

In order to demonstrate that the single aluminum atom contact realizes only a single dominant channel, we analyze the subgap current in the corresponding spectra \cite{cuevas_hamiltonian_1996}. Following previous findings  \cite{scheer_signature_1998,scheer_conduction_1997,cuevas_microscopic_1998}, we expect for the situtation of a superconducting contact made of an Al tip and an Al sample, that multiple Andreev reflections provide the most direct and most straightforward way of determining the the mesoscopic pin code. Experimental data for the current-voltage characteristics of the single-atom contact for different tip-sample distances are shown in Fig.\ \ref{fig:mars}(a). The tip-sample distance decreases from the dark blue spectrum to the yellow spectrum as the normal state conductance increases. We distinguish between the low voltage regime with a peak-like structure (Josephson regime, blue shaded, below $70\,\upmu$V) and the subgap MAR-regime with step-like structures. It is this latter regime that we explore first in order to fix the parameters that determine the physics of the low-voltage Josephson current.

We start with a differential conductance spectrum at a small normal conductance setpoint sufficiently away from the subgap domain, see Fig.\ \ref{fig:topo}(d). By fitting the density of states to the experimental data with the Bardeen-Cooper-Schrieffer (BCS) model of tip and sample, we find the values of the gap parameters as $\Delta_\textrm{tip}=180\,\upmu$eV and $\Delta_\textrm{sample}=180\,\upmu$eV. These are then fed into a standard MAR model \cite{cuevas_hamiltonian_1996,cuevas_dc_2004,cuevas_full_2003,averin_ac_1995} that extracts the mesoscopic pin code by assuming independent transmission channels to capture the experimental data at a given voltage (for details see the Supporting Information \cite{supinf}). The fits are shown in Fig.\ \ref{fig:mars}(a) as thinner lines with darker color superimposed on the data.  The fits accurately describe the MARs in the subgap regime ($<360\,\upmu$V) with small discrepancies only appearing at the onset of MAR steps at larger tip-sample distances. We attribute them to inelastic processes in the electron-hole tunneling which is not included in the modeling \cite{ast_sensing_2016} and requires an extended description accounting for features known from dynamical Coulomb blockade \cite{yeyati_dynamical_2005}.

In the mesoscopic pin code analysis, we use three transmission channels (cf.\ \cite{scheer_signature_1998,cuevas_evolution_1998}) and find that a single channel dominates the others by at least one order of magnitude for all tip-sample distances measured (see Fig.\ \ref{fig:mars}(b)). Thus, we conclude that in our STM set-up a single-atom contact between Al tip and sample with a single transmission channel is realized experimentally to very good approximation. With the transport parameters fixed, we can turn to the low-voltage regime (see Fig.\ \ref{fig:mars}(a)) to explore the Josephson effect in a rather unconventional domain, where charge transfer through a single channel with tunable transmission meets dynamical Coulomb blockade physics.

\begin{figure}
\centerline{\includegraphics[width = 1.1\columnwidth]{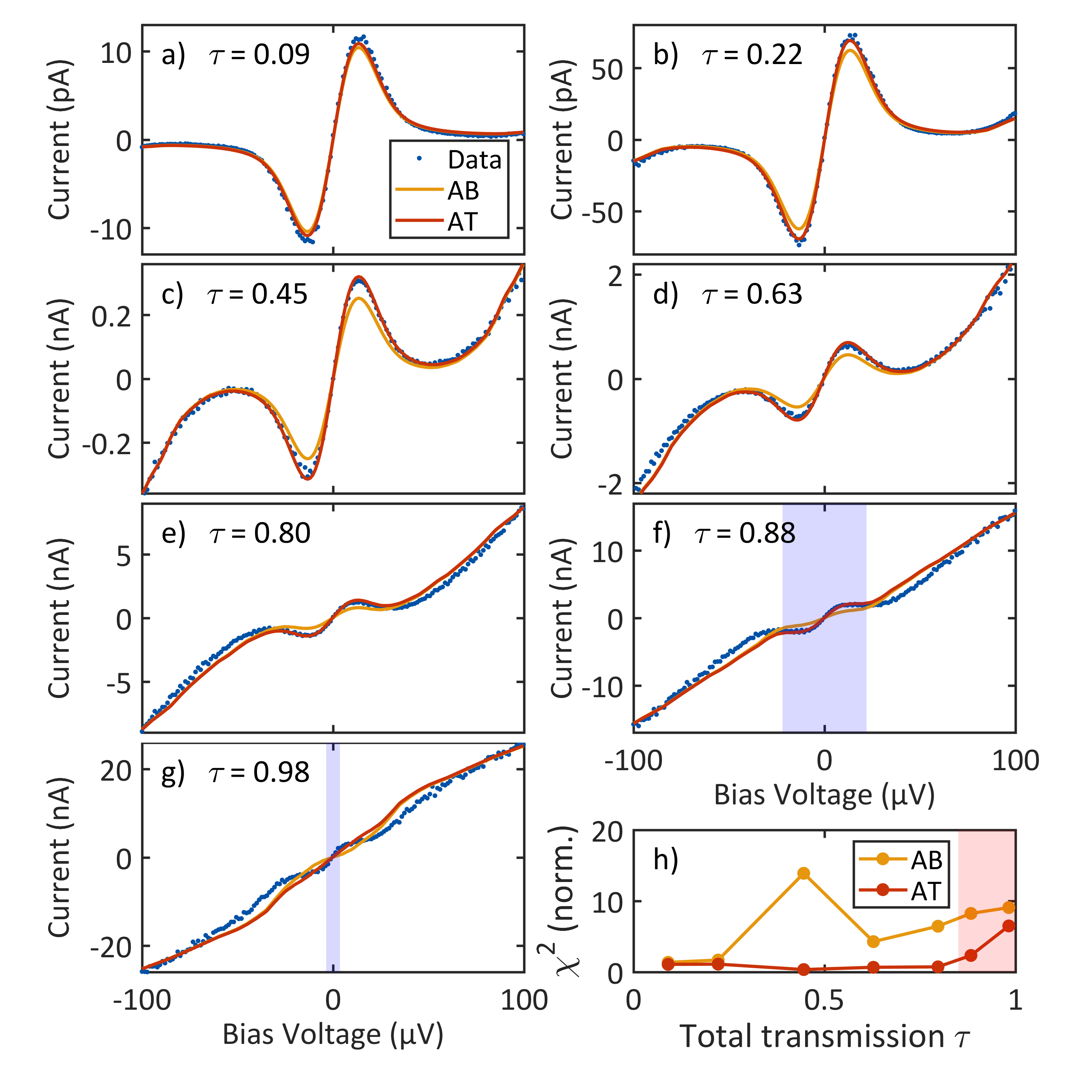}}
\caption{a)-g) Josephson spectra at different transmissions (the total transmission $\tau$ is indicated in each panel). The calculated spectra have essentially no free fit parameter, except for the first spectrum at lowest transmission. A clear deviation between the Ambegaokar-Baratoff approach (AB) and the full Andreev bound state relation at arbitrary transmission (AT) can be seen at higher transmission. For the transmissions in f) and g), non-adiabatic processes become significant at higher bias voltages, such that our model is only applicable within the blue shaded areas. Panel h) shows the $\chi^2$-values of the fits. The arbitrary transmission model yields low $\chi^2$-values throughout (indicating good agreement), except in the red shaded region, where non-adiabatic processes become significant.} \label{fig:jospanels}
\end{figure}

Figure \ref{fig:jospanels} shows Josephson spectra for different values of transmission of the single channel contact ranging from weak tunneling to nearly maximal transmission. The pronounced Josephson peak visible in the weak tunneling regime becomes washed out as the transmission increases, finally ending up masked by the background created by MARs. The Josephson peak arises due to inelastic Cooper pair tunneling, with broadening determined by the interaction with the environment. In our low-temperature STM experiment the granularity of charge transport determines the nature of these interactions and the observed peak is a manifestation of dynamical Coulomb-blockade. The interplay between (Josephson) tunneling and electromagnetic degrees of freedom of the environment can be described by $P(E)$-theory \cite{ingold_cooper-pair_1994,devoret_effect_1990,averin_coulomb_1990}. The current-voltage relation is
\begin{equation}
I(V) = \frac{2\pi}{\hbar}\left(\frac{E_\mathrm{J}}{2}\right)^2(2e)\left[ P(2eV)-P(-2eV)\right]
\label{eq:IVmulti}
\end{equation}
where the Josephson energy $E_\mathrm{J}$ is given by the AB formula
\begin{equation}
E_\mathrm{J} = \frac{\Delta}{4} \sum_i \tau_i = \frac{\Delta}{4} \frac{G_\mathrm{N}}{G_0}\,
\end{equation}
and $G_\text{N}$ is the total normal state conductance. The $P(E)$-function describes the probability density for exchanging energy $E$ with the environment. Here, the energy $E$ is given by the kinetic energy of the tunneling Cooper pair $2eV$, where $V$ is the applied junction bias voltage. This approach describes junctions with arbitrary many channels, as long as the transmission of each individual channel is small, $\tau_i \ll 1$.

The above theory for single-Cooper pair tunneling underestimates the Josephson peak at higher transmission (yellow line in Fig.\ \ref{fig:jospanels}), as compared to what is observed in our experimental data (blue dots in Fig.\ \ref{fig:jospanels}). It turns out that, at higher transmission, the nonlinear dependence of the energy on the transmission as well as to a smaller extent the tunneling of multiple Cooper pairs within a tunneling event are non-negligible and need to be taken into account. They can be traced back to the non-sinusoidal energy-phase relation expected for transparent single-channel contacts. In the following, we propose a simple extension of existing $P(E)$-theories that describes the \emph{dynamical Coulomb blockade regime of single and multiple Cooper pair transfer processes}.

The starting point is the ABS energy-phase relation for a single channel of arbitrary transmission \cite{beenakker_universal_1991,della_rocca_measurement_2007}
\begin{equation}
    E^{\pm}(\phi)=\pm\Delta\sqrt{1-\tau\sin^2{\left(\phi/2\right)}}
    \label{eq:abs}
\end{equation}
where the index $\pm$ labels the states between $-\Delta$ and $+\Delta$ below ($-$) and above ($+$) the Fermi level.
Focussing on the lower ($-$) Andreev state (i.\ e., assuming low temperature and the adiabatic limit, see below), we can express the energy as
\begin{equation}
    E(\phi)=\sum_{m=-\infty}^{+\infty}E_{m}e^{im\phi}
\end{equation}
with the coefficients $E_m$ given by
\begin{equation}
    E_{m}=-\Delta\sum_{k=|m|}^{+\infty}\binom{1/2}{k}\binom{2k}{k+m}(-1)^{m+k}(\tau/4)^{k}.
    \label{eq:coefficients}
\end{equation}

In the spirit of $P(E)$-theory, we replace the phase $\phi$ in the energy-phase relation by an operator. The phase acquires significant fluctuations in the dynamical Coulomb-blockade regime, where the number of transferred charges is a well-defined quantity. In the resulting Hamiltonian, a perturbative treatment is applied to the operators $e^{i m \phi}$ that induce the translation of $m$ Cooper pairs. Instead of the single $P(E)$-function in the tunnel limit describing inelastic single Cooper pair tunneling, each $m$-Cooper pair tunneling process involves a new $P_m(E)$-function that gives the probability of energy exchange $2meV$ (see Supplemental Material for details \cite{supinf}):
\begin{equation}
P_{m}(E)=\int_{-\infty}^{+\infty}\frac{\mathrm{d}t}{2\pi\hbar}e^{m^{2}J(t)+iEt/\hbar}
\end{equation}
and we find a Josephson current for the single channel case,
\begin{equation}
I(V)= \frac{2\pi}{\hbar}\sum_{m=1}^{+\infty}
\left|E_m\right|^2 (2me)\left[ P_m(2meV)-P_m(-2meV)\right]\,,
\label{eq:IVsingle}
\end{equation}
which depends on the $P_m(E)$ functions at the energies $\pm2meV$.
Note, that in the tunnel limit, $\tau \ll1$, of Eq.~\eqref{eq:coefficients} the coefficient $|E_{1}|=\Delta \tau/8=E_\text{J}/2$ dominates and Eq.~\eqref{eq:IVsingle} reduces to Eq.~\eqref{eq:IVmulti}.

The results of the extended DCB theory (red lines in Fig.\ \ref{fig:jospanels}) from Eq.~\eqref{eq:IVsingle} are compared to the experimental data and with the conventional DCB from Eq.~\eqref{eq:IVmulti}. In Fig.\ \ref{fig:jospanels}(h), the $\chi^2$-values for the calculated curves are plotted as function of total transmission. The lower $\chi^2$-values for the arbitrary transmission (AT) model indicate a much better agreement compared to the AB model (details of the $\chi^2$ calculation can be found in the Supporting Information \cite{supinf}). Crucially, both theoretical calculations do not involve any free fit parameters, but rely only on the mesoscopic pin code known from the MAR analysis, gap parameters for tip and sample obtained from the quasiparticle spectrum at low conductance, and on the tunnel junction parameters entering the $P_{(m)}(E)$-function(s) (see Supporting Information \cite{supinf}) determined by the Josephson-spectrum at lowest transmission \cite{fn}.
Without introducing additional parameters or assumptions, the I-V given by Eq.~\eqref{eq:IVsingle} clearly improves upon the conventional DCB result. We find good agreement with the experimental data over the whole voltage range of the Josephson peak up to $\tau\sim 0.7$. For larger $\tau$ this voltage range shrinks until no discernible agreement can be claimed at the highest transmission, $1-\tau \ll 1$.

\begin{figure}
\centerline{\includegraphics[width = \columnwidth]{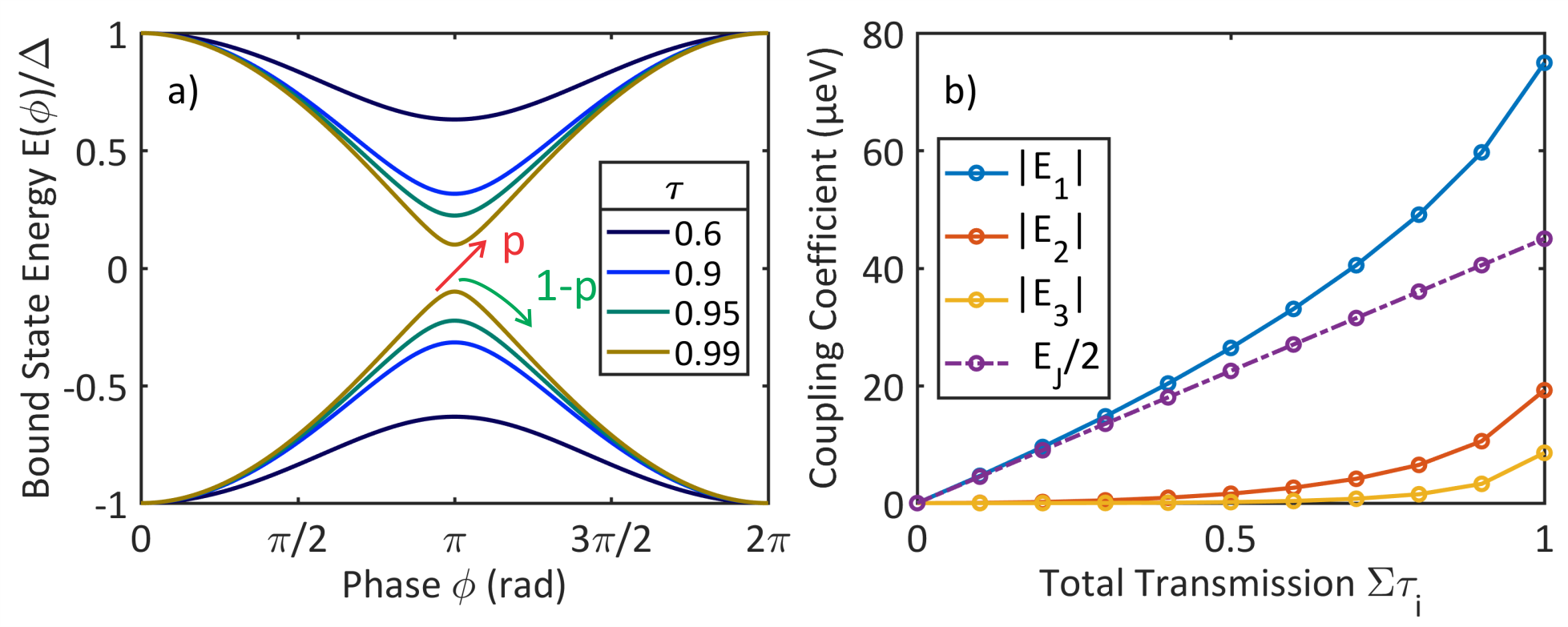}}
\caption{a) Andreev bound state relation for high transmissions. As the gap closes, the probability for transitions between branches (non-adiabatic processes) becomes more likely. b) Absolute value of the coupling coefficients $|E_m|$ at different transmissions in comparison to the coupling coefficient of the linear Ambegaokar-Baratoff model ($E_\text{J}/2$).} \label{fig:couplingcoeffs}
\end{figure}

The level to which our extension of DCB-theory reproduces the STM measurements is fully consistent with expectations:
(i) The improved agreement at high transmission can be attributed to the dependence of coefficients $E_m$ on the channel transmission $\tau$. These coefficients $E_m$ of the extended theory play a similar role to $E_\text{J}$ in Eq.\ \eqref{eq:IVmulti}. The comparison between $E_m$ for ($m = 1, 2, 3$) and $E_\text{J}/2$ is plotted in Fig. 4(b) showing a significant increase of $E_1$ compared to $E_\text{J}/2$ for transmissions $\tau\gtrsim 0.2$. At higher transmission the probability for multiple Cooper pair transfers increases, such that a (small) part of the current is due to the energy exchange of multiple Cooper pairs with the environment. Distinguishing the contributions of processes with different $m$ is not possible in our experiment. In the future, this could be achieved using a designed environment with sufficiently narrow resonances, such that specific $m$-Cooper pair process can be enhanced via the $P_m(E)$-functions in Eq.\ \eqref{eq:IVsingle}.
(ii) As a low-order perturbative approach any $P(E)$-type approach is bound to fail, if tunneling becomes too frequent for the environment to relax between consecutive tunneling events. A simple test (see Supporting Information \cite{supinf}) suggests that the assumption of sequential, independent tunneling events inherent to the rate-picture of Eqs.\ \eqref{eq:IVmulti} and \eqref{eq:IVsingle} breaks down at about $\tau \sim 0.9$.
(iii) Even without environmental effects, however, the large transmission limit, $1-\tau \ll 1$, is challenging, since we can no longer separate a large-voltage regime of MAR from the low-voltage region of Josephson-tunneling, cf.\ Fig.\ \ref{fig:mars}(a), as high-order Andreev reflections scale with high powers of the transmission and are no longer suppressed. Within the equilibrium picture of ABS, $E^\pm(\phi)$ in Eq.~\eqref{eq:abs}, with $\phi$ being a classical variable, in the low-bias regime, the dissipative current can be understood in terms of Landau-Zener transitions between the ABSs, cf.\ Fig.~\ref{fig:couplingcoeffs}(a) \cite{chauvin_crossover_2007}. While well in the dynamical Coulomb blockade regime of large quantum fluctuations of the phase, we can, nonetheless, use the classical phase picture and exploit the Landau-Zener transition probability, $p=\exp{\left[-\pi(1-\tau)\Delta/eV\right]}$, to estimate a threshold voltage, $V = (1-\tau)\Delta$, where non-adiabatic transitions become non-negligible. This voltage, diminishing as the transmission approaches unity, fits with the observed range of validity in Fig.~\ref{fig:jospanels}. For instance, for $\tau= 0.88$ [Fig.~\ref{fig:jospanels}(g)], we expect and observe the adiabatic model to fail outside of the region $-22\mu V \le V \le 22\mu V$ shaded in blue.

Further theoretical and experimental investigation of the large transmission regime will advance a more complete understanding, complementing the elaborate, self-consistent theory for the case of thermal phase fluctuations \cite{chauvin_crossover_2007}, and can also clarify the impact of a renormalization of the charging energy at stronger tunneling \cite{jezouin_controlling_2016}.

We have demonstrated the single channel Josephson effect in the STM from an atomic contact at arbitrary transmission up to the quantum of conductance and in presence of dynamical Coulomb blockade. The Josephson current in this regime can be very well modeled by the energy-phase relation of the full Andreev bound state projected onto the charge transfer operators for single and multiple Cooper pair tunneling. We find excellent agreement between theory and experiment with no free parameters as each parameter has been determined independently. Concerning few channel junctions, we believe, it is crucial to consider individual transport channels and their transmission separately instead of using the total conductance and the multi-channel approximation of the AB formula. Deviations can already be discerned at channel transmissions as low as $\tau=0.1$.

These findings are an important step towards a detailed understanding of the dynamical Coulomb blockade regime in the Josephson effect measured by the STM. This is urgently needed, as the Josephson effect becomes important as a tool to extract local information about the superconducting properties of the sample. For instance, when scanning across a magnetic adatom that induces Yu-Shiba-Rusinov states and locally reduces the order parameter of the substrate \cite{salkola_spectral_1997,flatte_local_1997-1}, the inevitable changes in the number of channels and their transmission have to be considered.

We gratefully acknowledge fruitful discussions with Berthold Jäck, and Elke Scheer. Funding from the European Research Council for the Consolidator Grant ABSOLUTESPIN, from the Spanish MINECO (Grant No.\ FIS2014-55486-P, FIS2017-84057-P, and FIS2017-84860-R), from the ``Mar\'{\i}a de Maeztu'' Programme for Units of Excellence in R\&D (MDM-2014-0377), from the Zeiss Foundation, from the DFG through AN336/11-1, and from the IQST is also gratefully acknowledged.

\strut
\onecolumngrid
\newpage
\begin{center}
\textbf{\large Supplementary Information}
\strut
\vspace{1em}
\end{center}
\setcounter{figure}{0}
\setcounter{table}{0}
\renewcommand{\thefigure}{S\arabic{figure}}
\renewcommand{\thetable}{S\Roman{table}}
\twocolumngrid
\section{Tip and Sample Preparation}

The experiments were carried out in a scanning tunneling microscope (STM) operating at a base temperature of 15\,mK \cite{si_assig_10_2013}. For the sample, we use an Al(100) single crystal. To obtain a clean Al(100) surface, the sample was cleaned by multiple cycles of Ar sputtering and subsequent annealing. The tip material was a polycrystalline Al wire, which was cut in air and prepared in ultrahigh vacuum by sputtering and field emission. Single Al atoms were pulled from the surface by the tip and then placed on the surface to realize a single atomic contact. We demonstrate that the tip apex remains unaffected by the manipulation through imaging of the reference impurity (sombrero shape in Fig.\ \ref{fig:topo}(a) and (b) of the main text). Both images of the reference impurity look identical before and after the manipulation.

\section{Extracting the Mesoscopic Pin Code from the subgap current}

\begin{figure}
\centerline{\includegraphics[width = \columnwidth]{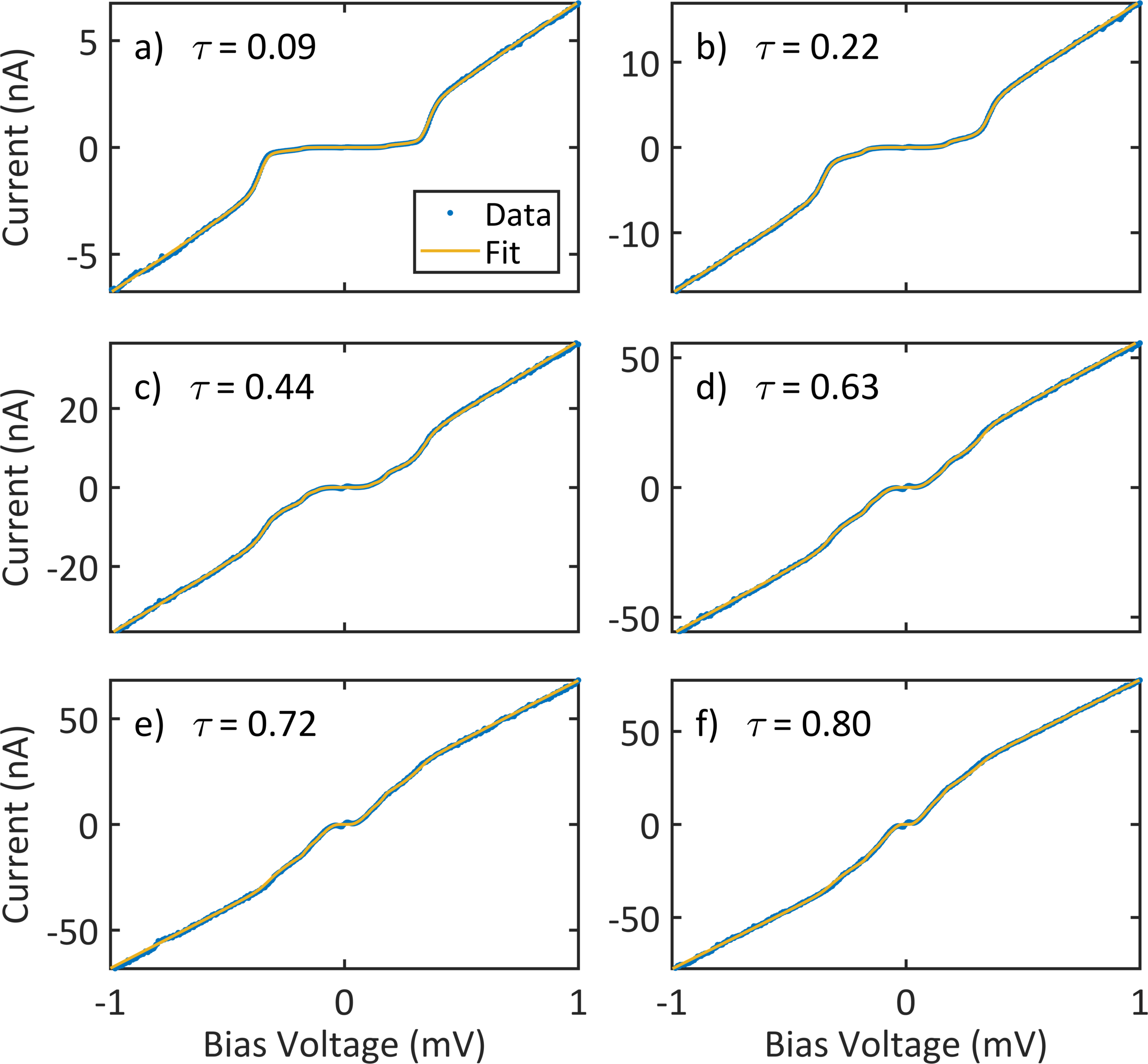}}
\caption{Linear plot of the experimental multiple Andreev reflection spectra along with their corresponding fits.} \label{fig:marlin}
\end{figure}

To extract the number of channels and their transmission, we exploit the multiple Andreev reflections that are measured at different normal state conductances. For the fits, we use the model of multiple Andreev reflections outlined in Ref.\ \cite{Cuevas} treating multiple channels as independent of each other and adding them for the total spectrum. As has been shown before \cite{Ast}, the aluminum spectra cannot be fitted accurately at low temperatures without considering the spectral broadening due to the interaction with the environment ($P(E)$-theory). However, the models considered in the context of multiple Andreev reflections do not include the environmental interactions to the degree needed here. In addition, the broadening cannot be modelled by an effective Dynes parameter $\Gamma$ either \cite{Dynes}. To remedy this shortcoming, we approximate the $P(E)$-function by a Gaussian $P^{\ast}(E)$, which includes an effective capacitive voltage noise. In this way, the broadening is symmetric and we can include the influence of the environment at least phenomenologically to lowest order by convolving one of the leads' density of states with this $P^{\ast}(E)$-function before calculating the Andreev reflections. (For the later analysis of the Josephson spectra, we employ the full $P(E)$-function.) To determine the fitting parameters, we fit a spectrum at low conductance that does not show any Andreev reflections nor any Josephson effect. We calculate the differential conductance $dI/dV$ from the tunneling current
\begin{equation}
I(V) = e\left(\vec{\mathit\Gamma}(V)-\cev{\mathit\Gamma}(V)\right),
\label{eq:iv}
\end{equation}
with the tunneling probability from tip to sample
\begin{widetext}
\begin{equation}
\vec{\mathit\Gamma}(V)=\frac{1}{e^2R_\text{T}}\int\limits^{\infty}_{-\infty}\int\limits^{\infty}_{-\infty}dEdE'\rho_\text{t}(E)\rho_\text{s}(E'+eV)
f(E)[1-f(E'+eV)]P^{\ast}(E-E').
\label{eq:tunprob}
\end{equation}
\end{widetext}
The other tunneling direction $\cev{\mathit\Gamma}(V)$ from sample to tip can be obtained by exchanging electrons and holes in Eq.\ \eqref{eq:tunprob}. Here, $R_\text{T}$ is the tunneling resistance, $f(E)=1/(1+\exp(E/k_\text{B}T))$ is the Fermi function, and $\rho_\text{t}$, $\rho_\text{s}$ are the densities of states of tip and sample, respectively. Here, we replace the full $P(E)$-function by a Gaussian modeling an effective capacitive noise:
\begin{equation}
P^{\ast}(E)=\frac{1}{\sqrt{4\pi E_\text{C} k_\text{B}T}}\exp\left[-\frac{E^2}{4E_\text{C}k_\text{B}T}\right]
\end{equation}
where $E_C=Q^2/2C_\text{J}$ is the (effective) charging energy for Cooper pairs ($Q=2e$). We use the same effective capacitance $C_\text{J}=21.7\,$fF and effective temperature $T=200\,$mK as for the modeling of the Josephson effect, which does not introduce any new parameters (see below). The densities of states of the aluminum tip and the aluminum sample are given by the simple Bardeen-Cooper-Schrieffer (BCS) expression:
\begin{equation}
    \rho_\text{t,s}(\omega) = \Re\left[\frac{\omega+i\Gamma_\text{t,s}}{\sqrt{(\omega+i\Gamma_\text{t,s})^2-\Delta_\text{t,s}^2}}\right],
\end{equation}
where the order parameters for both tip and sample are the same $\Delta_\text{t,s}=180\,\upmu$eV and the Dynes parameter $\Gamma_\text{t,s}=0.01\,\upmu$eV is only non-zero for numerical purposes.

In order to fit the number of channels and their transmissions to the experimental data, we calculate a series of Andreev reflection spectra from the fit parameters for different transmissions. Assuming independent channels, we fit linear combinations of these spectra to the experimental data using a statistical search algorithm (Matlab).

We should point out, that the result of having single channel transmission is quite robust with respect to details of the fit and the modeling, i.\ e.\ whether we use the symmetric $P^{\ast}(E)$-function or a phenomenological broadening parameter. This is because the number of channels and their conductance do not only modify the subgap structure of the spectrum, but they also have a sizeable effect on the normal conducting part of the spectrum, i.\ e.\ at voltages $|V|>\Delta_\text{t}+\Delta_\text{s}$. The so-called excess current $I_\text{exc}$ \cite{Cuevas} is a constant (voltage independent) current contribution to the normal conducting part of the spectrum. The excess current $I_\text{exc}$ increases monotonically with the channel transmission and is quite independent of the details of the modeling as well as of any broadening mechanisms. It is easy to see that the excess current $I_\text{exc}$ contribution becomes maximal in the case of a single channel contact. The excess current $I_\text{exc}$ is naturally included in the Andreev calculations and as such the whole spectrum inside and outside of the gap becomes an important indicator for single channel tunneling.

\section{Calculating the Josephson Current within \textit{P(E)}-Theory at Arbitrary Transmission}

The Josephson current in the STM at mK temperatures and \textit{low} transmission, where only single Cooper-pair processes play a role, has previously successfully been modeled on the basis of the simple sinusoidal energy-phase relation and the standard $P(E)$-function for single Cooper pairs (cf. Eq.~(1) in the main text). To calculate the $I(V)$ characteristics of the single-channel Josephson effect at \textit{arbitary} transmission, we start from the full energy-phase relation $E^{\pm}(\phi)$ of the Andreev bound states (cf. Eq.~(3) in the main text), which includes multiple Cooper-pair processes. Since we only consider non-adiabatic processes in this model, we focus on the lower branch ($-$ label) and omit the sign index in the following. We transform this energy-phase relation from phase space to charge space by a Fourier transform, $E_{\phi}=\sum^{\infty}_{m=-\infty}E_{m}e^{im\phi}$ (cf. Eq.~(4) and (5) in the main text), in order to include the interaction of multiple Cooper-pair transfers with the surrounding environment. Following the standard procedure of calculating tunneling currents within the framework of $P(E)$-theory (see, e.g., Refs.~\cite{Devoret1990,Ingold,Ingold1994}), the tunneling rate associated with a forward tunneling process of $m$ Cooper pairs through a single channel is given by
\begin{equation}
\Gamma_{m}=\frac{2\pi}{\hbar}|E_{m}|^{2}\sum_{R_i,R_f}|\braket{R_{f}|e^{-im\phi}|R_{i}}|^{2}P_{\beta}(R_{i})\delta(E_{R_i}-E_{R_f}).
\end{equation}
The corresponding backwards tunneling rates can be obtained by replacing $m$ by $-m$. This is essentially a golden rule rate with $E_{m}e^{-im\phi}$ as the perturbation. Here, we sum over all initial reservoir states $R_i$ with energies $E_{R_i}$ weighted by the probability $P_{\beta}(R_i)$ to find these states at the inverse temperature $\beta=1/(k_{\rm B}T)$ and over all final states $R_f$ with energies $E_{R_f}$. Performing the trace over the environmental degrees of freedom and exploiting the generalized Wick theorem, we finally arrive at
\begin{equation}
\Gamma_{m}=\frac{2\pi}{\hbar}|E_{m}|^{2}P_{m}(m2eV).
\end{equation}
Here, we have introduced the generalized $P(E)$-function
\begin{equation}
P_{m}(E)=\int_{-\infty}^{+\infty}\frac{\mathrm{d}t}{2\pi\hbar}e^{m^{2}J(t)+iEt/\hbar}
\end{equation}
with the phase correlation function
\begin{equation}
J(t)=\braket{[\tilde{\phi}(t)-\tilde{\phi}(0)]\tilde{\phi}(0)},
\end{equation}
accounting for Cooper-pair--phase fluctuations $\tilde{\phi}=\phi-2eVt/\hbar$ around the mean value determined by the external voltage. For $m=\pm1$, we recover again the standard $P(E)$-function used in the low-transmission limit for single Cooper-pair processes. Summing up all possible processes (forward and backward tunneling of one, two, up to infinitely many Cooper pairs), the current contribution of the lower Andreev bound state can be written as
\begin{align}
I(V)&=\sum_{m=-\infty}^{+\infty}(2me)\Gamma_{m}\\
&=\frac{2\pi}{\hbar} \sum_{m=1}^{+\infty}|E_{m}|^{2}(2me)\left[P_{m}(m2eV)-P_{m}(-m2eV)\right].
\end{align}

\section{Validity of the \textit{P(E)}-theory}

\begin{figure}
\centerline{\includegraphics[width = 0.9\columnwidth]{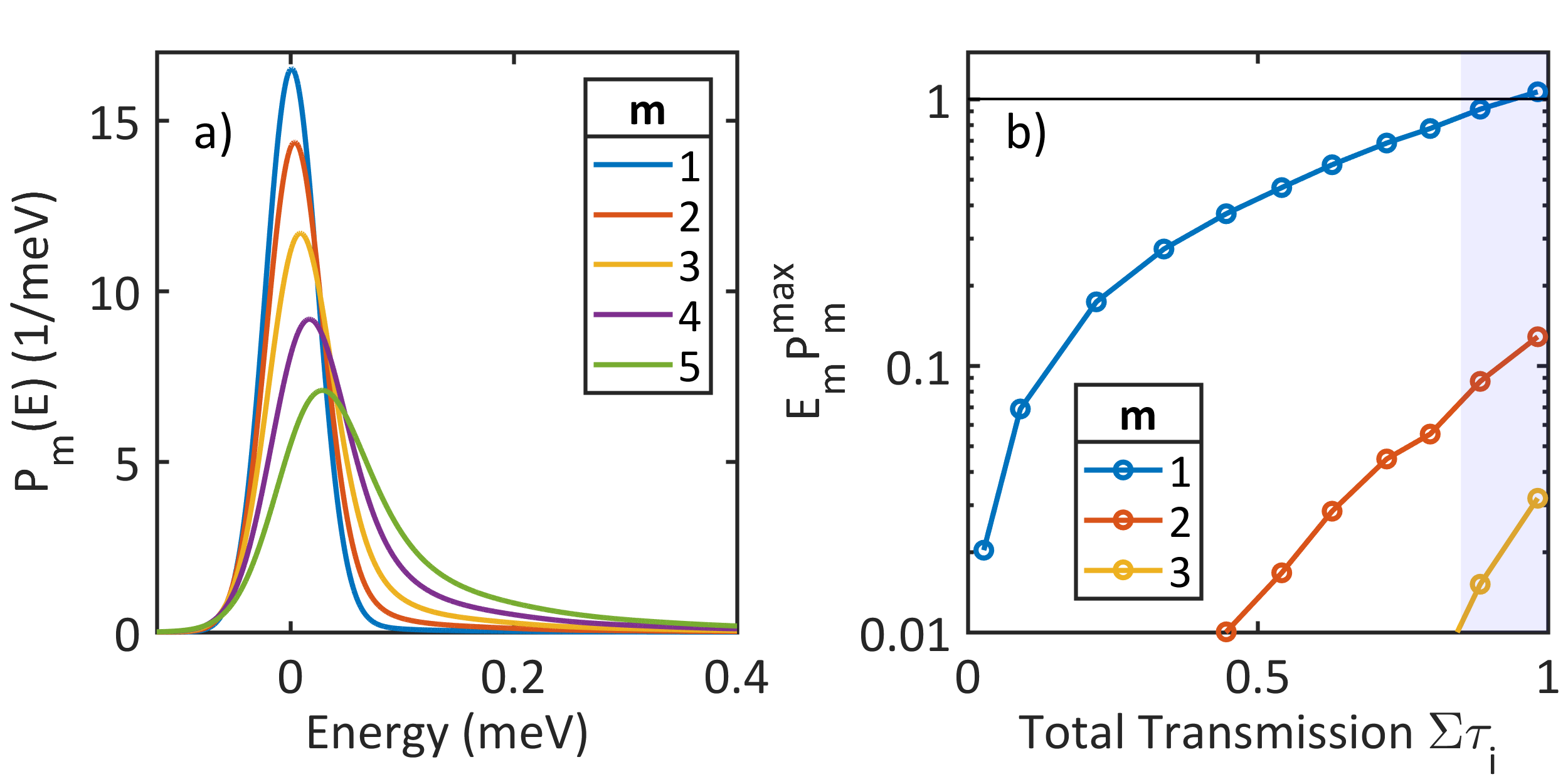}}
\caption{a) Plots of the $P(E)$-function for $m$ transferred Cooper pairs. For higher orders of $m$, the weight shifts to higher energies. b) Check for the applicability of $P(E)$-theory for single and multiple Cooper pair transfers. For all transmissions except the highest ($\tau>0.9$) the product of coupling constant $E_\text{m}$ with the maximum value of the $P_\text{m}(E)$-function $P_\text{m}^\text{max}$ is less than one, so that $P(E)$-theory is applicable. The blue shaded area shows the region, where non-adiabatic processes start dominating the spectrum.} \label{fig:validity}
\end{figure}

To test for the applicability of $P(E)$-theory in this context, we calculate the product of the coupling constant with the $P(E)$-function \cite{Ingold}. In the tunneling regime, this condition of validity is typically written as:
\begin{equation}
    E_\text{J}P(E)\ll 1
\end{equation}
The $P(E)$-functions used in this analysis are plotted in Fig.\ \ref{fig:validity}(a). The parameters are given in the next section. The different curves are for multiple Cooper pair transfers up to $m=5$. In order to simplify the analysis, we just use the maximum value of the $P(E)$-function $P^\text{max}$, which in our scenario is typically near $E=0$. Here, we consider sequential tunneling of single and multiple Cooper pairs, so we will calculate the product for each value of $m$ separately. Therefore, the coefficient $E_\text{m}$ takes the role of the coupling constant and the maximum of $P_\text{m}(E)$ is $P_\text{m}^\text{max}$, such that the condition reads as:
\begin{equation}
    |E_m|P_m^\text{max}\ll 1
\end{equation}
The values for different Cooper pair transfers up to $m=3$ are plotted in Fig.\ \ref{fig:validity}(b). We can see that all values are less than one for all transmissions except the highest ones. Here, the validity condition is broken for single Cooper pair transfers at transmissions $\tau>0.9$. This is in agreement with our previous analysis of the validity of $P(E)$-theory in the DCB regime of the STM \cite{Jack}. As we have shown in the main text that at these transmission values also non-adiabatic processes start to dominate, which are not directly captured in the theoretical model, breaking the applicability condition at these high transmission values is of minor relevance.

At the highest transmission of $\tau = 0.98$, as shown in Fig.\ 3(h) in the main text, the Josephson effect cannot be modeled adequately anymore by our models. This is due to the non-adiabatic processes becoming dominant as discussed in the main text as well as a breakdown of the applicability condition for $P(E)$-theory as shown in Fig.\ \ref{fig:validity}.

\section{Calculating the $\chi^2$-values}

We define the dimensionless $\chi^2$-values as
\begin{equation}
    \chi^2 = \sum\limits_i\frac{(y_i^\text{data}-y_i^\text{model})^2}{(y_i^\text{data})^2}.
\end{equation}
where $i$ indexes all data points in the data set, $y_i^\text{data}$ is a data point, and $y_i^\text{model}$ is the calculated value corresponding to the data point. The $\chi^2$-value is thus a dimensionless, normalized quantity indicating the deviation of the calculation from the data. The lower the $\chi^2$-value, the better the agreement. The $\chi^2$-values have been calculated for a voltage interval $|V|\le30\,\upmu$eV.

\section{Modeling the Josephson Spectra}

The Josephson effect is modeled with the full $P(E)$-function. We cannot employ the simplified $P^{\ast}(E)$-function used in the analysis of the Andreev spectra, because they are symmetric and would yield zero Josephson current. In order to extract the relevant parameters, we fit a Josephson spectrum at very small conductances, where we can assume little difference between the Ambegaokar-Baratoff approach and the full Andreev bound state relation. The fit function is based on the equation:
\begin{equation}
    I(V)=\frac{\pi e E_\text{J}^2}{\hbar}\left[P(2eV)-P(-2eV)\right],
\end{equation}
where $E_\text{J}$ is the Josephson energy. The parameters are for the junction capacitance $C_\text{J}=21\,$fF and for the effective temperature $T=200\,$mK. For the environmental impedance $Z(\omega)$, we use the modified transmission line impedance introduced in Ref.\ \cite{Jack}, which describes the impedance of the surrounding vacuum as well as the resonances of the tip acting as a monopole antenna. The parameters for the modified transmission line impedance are the resonance energy $\omega_\text{R}=70\,\upmu$eV, the effective damping parameter $\alpha=0.9$, and the environmental dc resistance $R_\text{env}=377\,\upOmega$ \cite{Jack}.


\begin{thebibliography}{10}
\bibitem{muller_experimental_1992}
Muller, C.~J., van Ruitenbeek, J.~M., and de Jongh, L.~J.
\newblock {\em Physica C} {\bf 191}, 485 (1992).

\bibitem{scheer_signature_1998}
Scheer, E., Agra\"it, N., Cuevas, J.~C., Yeyati, A.~L., Ludoph, B.,
  Martín-Rodero, A., Bollinger, G.~R., van Ruitenbeek, J.~M., and Urbina, C.
\newblock {\em Nature}{ \bf 394}, 154 (1998).

\bibitem{scheer_conduction_1997}
Scheer, E., Joyez, P., Esteve, D., Urbina, C., and Devoret, M.~H.
\newblock {\em Phys. Rev. Lett.}{ \bf 78}(18), 3535 May  (1997).

\bibitem{chauvin_crossover_2007}
Chauvin, M., vom Stein, P., Esteve, D., Urbina, C., Cuevas, J.~C., and Yeyati,
  A.~L.
\newblock {\em Phys. Rev. Lett.}{ \bf 99}, 067008 (2007).

\bibitem{della_rocca_measurement_2007}
Della~Rocca, M.~L., Chauvin, M., Huard, B., Pothier, H., Esteve, D., and
  Urbina, C.
\newblock {\em Phys. Rev. Lett.}{ \bf 99}, 127005 (2007).

\bibitem{ambegaokar_tunneling_1963}
Ambegaokar, V. and Baratoff, A.
\newblock {\em Phys. Rev. Lett.}{ \bf 10}, 486 (1963).

\bibitem{beenakker_universal_1991}
Beenakker, C. W.~J.
\newblock {\em Phys. Rev. Lett.}{ \bf 67}, 3836 (1991).

\bibitem{ingold_cooper-pair_1994}
Ingold, G., Grabert, H., and Eberhardt, U.
\newblock {\em Phys. Rev. B}{ \bf 50}, 395 (1994).

\bibitem{jack_nanoscale_2015}
J\"ack, B., Eltschka, M., Assig, M., Hardock, A., Etzkorn, M., Ast, C.~R., and
  Kern, K.
\newblock {\em Appl. Phys. Lett.}{ \bf 106}, 013109 (2015).

\bibitem{ast_sensing_2016}
Ast, C.~R., J{\"a}ck, B., Senkpiel, J., Eltschka, M., Etzkorn, M., Ankerhold,
  J., and Kern, K.
\newblock {\em Nature Commun.}{ \bf 7}, 13009 (2016).

\bibitem{jack_quantum_2017}
Jäck, B., Senkpiel, J., Etzkorn, M., Ankerhold, J., Ast, C.~R., and Kern, K.
\newblock {\em Phys. Rev. Lett.}{ \bf 119}, 147702 (2017).

\bibitem{devoret_effect_1990}
Devoret, M.~H., Esteve, D., Grabert, H., Ingold, G., Pothier, H., and Urbina,
  C.
\newblock {\em Phys. Rev. Lett.}{ \bf 64}, 1824 (1990).

\bibitem{averin_incoherent_1990}
Averin, D., Nazarov, Y., and Odintsov, A.
\newblock {\em Physica B: Cond. Mat.}{ \bf 165-166}, 945 (1990).

\bibitem{cuevas_hamiltonian_1996}
Cuevas, J.~C., Martín-Rodero, A., and Yeyati, A.~L.
\newblock {\em Phys. Rev. B}{ \bf 54}, 7366 (1996).

\bibitem{jack_critical_2016}
J\"ack, B., Eltschka, M., Assig, M., Etzkorn, M., Ast, C.~R., and Kern, K.
\newblock {\em Phys. Rev. B}{ \bf 93}, 020504 (2016).

\bibitem{randeria_scanning_2016}
Randeria, M.~T., Feldman, B.~E., Drozdov, I.~K., and Yazdani, A.
\newblock {\em Phys. Rev. B}{ \bf 93}, 161115 (2016).

\bibitem{hamidian_detection_2016}
Hamidian, M.~H., Edkins, S.~D., Joo, S.~H., Kostin, A., Eisaki, H., Uchida, S.,
  Lawler, M.~J., Kim, E.-A., Mackenzie, A.~P., Fujita, K., Lee, J., and Davis,
  J. C.~S.
\newblock {\em Nature}{ \bf 532}, 343 (2016).

\bibitem{cuevas_microscopic_1998}
Cuevas, J.~C., Yeyati, A.~L., and Martín-Rodero, A.
\newblock {\em Phys. Rev. Lett.}{ \bf 80}, 1066 (1998).

\bibitem{cuevas_dc_2004}
Cuevas, J.~C. and Belzig, W.
\newblock {\em Phys. Rev. B}{ \bf 70}, 214512 (2004).

\bibitem{cuevas_full_2003}
Cuevas, J.~C. and Belzig, W.
\newblock {\em Phys. Rev. Lett.}{ \bf 91}, 187001 (2003).

\bibitem{averin_ac_1995}
Averin, D. and Bardas, A.
\newblock {\em Phys. Rev. Lett.}{ \bf 75}, 1831 (1995).

\bibitem{supinf}
see Supporting Information.

\bibitem{yeyati_dynamical_2005}
Levy~Yeyati, A., Cuevas, J.~C., and Martín-Rodero, A.
\newblock {\em Phys. Rev. Lett.}{ \bf 95}, 056804 (2005).

\bibitem{cuevas_evolution_1998}
Cuevas, J.~C., Levy~Yeyati, A., Martín-Rodero, A., Rubio~Bollinger, G.,
  Untiedt, C., and Agra\"it, N.
\newblock {\em Phys. Rev. Lett.}{ \bf 81}, 2990 (1998).

\bibitem{averin_coulomb_1990}
Averin, D. and Nazarov, Y.
\newblock {\em Physica B: Cond. Met.}{ \bf 162}, 309 (1990).

\bibitem{salkola_spectral_1997}
Salkola, M.~I., Balatsky, A.~V., and Schrieffer, J.~R.
\newblock {\em Phys. Rev. B}{ \bf 55}, 12648 (1997).

\bibitem{flatte_local_1997-1}
Flatt{\'e}, M.~E. and Byers, J.~M.
\newblock {\em Phys. Rev. Lett.}{ \bf 78}, 3761 (1997).

\bibitem{fn}
A ``background'' of Andreev reflections originating from extrapolating the
Andreev spectra in Fig.\ \ref{fig:mars}(a) into the low voltage range between
$\pm 100\,\upmu$V is also added. Such simple superposition of the Josephson effect
and the multiple Andreev reflections is valid as long as non-adiabatic processes are
negligible, which are not included in the model here as we will discuss later
\cite{chauvin_crossover_2007}.

\bibitem{jezouin_controlling_2016}
Jezouin, S., Iftikhar, Z., Anthore, A., Parmentier, F. D., Gennser, U., Cavanna, A., Ouerghi, A.,
Levkivskyi, I. P., Idrisov, E., Sukhorukov, E. V., Glazman, L. I., Pierre, F.
\newblock {\em Nature} {\bf 58}, 536 (2016).

\end{thebibliography}

\begin{thebibliography}{1}
\bibitem{si_assig_10_2013}
    M. Assig, \textit{et al.}, \textit{Rev. Sci. Instr.}, \textbf{84}, 033903 (2013).
\bibitem{Cuevas}
    J. C. Cuevas, A. Martin-Rodero, and A. Levy Yeyati, \textit{Phys. Rev. B}, \textbf{54}, 7366 (1996).
\bibitem{Ast}
    C.~R. Ast \textit{et al.}, {\em Nature Commun.}{ \bf 7}, 13009 (2016).
\bibitem{Dynes}
    R. C. Dynes, V. Narayanamurti, and J. P. Garno, \textit{Phys. Rev. Lett.}, \textbf{41}, 1509 (1978).
\bibitem{Ingold}
    G.~L. Ingold and Y.~V. Nazarov, in \textit{Single Charge Tunneling} edited by H. Grabert and M.H. Devoret,
    \textit{NATO ASI Series B} \textbf{294}, 21 (Plenum Press, New York, 1992).
\bibitem{Devoret1990} M.~H. Devoret, D. Esteve, H. Grabert, G.-L. Ingold, H. Pothier, and C. Urbina,
    Phys. Rev. Lett. \textbf{64}, 1824 (1990).
\bibitem{Ingold1994} G.-L. Ingold, H. Grabert, and U. Eberhardt, Phys. Rev. B \textbf{50}, 395 (1994).
\bibitem{Jack}
    B. Jäck \textit{et al.}, {\em Phys. Rev. B}{ \bf 93}, 020504 (2016).
\end{thebibliography}
\end{document}